\documentclass{aa}  

\usepackage{graphicx}
\usepackage{txfonts}
\usepackage{lipsum}
\usepackage{subcaption}
\usepackage{lscape}
\usepackage{placeins}
\usepackage{multirow}
\usepackage{hyperref}
\hypersetup{
    colorlinks=true,
    linkcolor=blue,
    filecolor=blue,      
    urlcolor=blue,
    citecolor=blue,
}

\begin{document}
   \title{Metal sign of a large-scale AGN feedback in cool-core cluster MACS J1931.8-2634}
    \author{Qinnan Zhu\inst{1}\corrauth{zqn23@mails.tsinghua.edu.cn}
        \and Junjie Mao\inst{1}\corrauth{jmao@tsinghua.edu.cn}
        \and Ping Zhou \inst{2,5}
        \and Yuanyuan Su \inst{3}
        \and Dan Hu \inst{4}
        }
    \institute{Department of Astronomy, Tsinghua University, Beijing 10084, China 
        \and School of Astronomy and Space Science, Nanjing University, Nanjing 210023, China 
        \and Department of Physics and Astronomy, University of Kentucky, 505 Rose Street, Lexington, KY, 40506, USA
        \and Yunnan Observatories, Chinese Academy of Sciences, Kunming 650216, China
        \and Key Laboratory of Modern Astronomy and Astrophysics, Nanjing University, Ministry of Education, Nanjing 210023, China
    }
   \date{}
 
  \abstract 
   {The spatial distribution of metals in the intracluster medium (ICM) is a sensitive tracer of the chemical and dynamical history of galaxy clusters. While most cool-core (CC) clusters exhibit a centrally peaked Fe abundance profile, several outliers show an anomalous central Fe drop, potentially associated with the AGN activities.}
   {We revisit the reported large-scale ($\sim100~{\rm kpc}$) central Fe drop in the massive CC cluster MACS J1931.8-2634 using new XMM-Newton observations. We aim to verify this feature and search for imprints of AGN feedback on the ICM metallicity distribution.} 
   {We analyzed $\sim170~\rm ks$ of new XMM-Newton observations and re-analyzed $\sim100~{\rm ks}$ archived Chandra observations. We derived radial and two-dimensional (2D) Fe abundance maps from CCD spectra. High-resolution RGS spectra were used to constrain the Ne/Fe abundance ratio to test the dust depletion scenario. Spectral fitting was performed in SPEX using an updated atomic database and both single- and multi-temperature collisional ionization equilibrium models.}
   {The previously reported central Fe drop is not confirmed in the radial profile from XMM-Newton. However, the 2D Fe distribution is clearly asymmetric: Fe-rich regions are elongated along the axis of the AGN cavities, extending beyond their immediate scale. The Ne/Fe ratio in the core is consistent with solar (Ne/Fe $= 1.03^{+0.25}_{-0.23}$), arguing against the dust depletion scenario.}
   {}
   \keywords{techniques: spectroscopic -- galaxies: clusters: individual:: MACS J1931.8-2634 -- X-rays: galaxies: clusters
   }
   \maketitle
   \nolinenumbers

\section{Introduction}

While the abundance of iron is commonly found to peak in the center of cool-core (CC) groups and clusters, several targets are observed with an unexpected central abundance drop at range of $2-50~{\rm kpc}$ (\citealp{2015MNRAS.447..417P}, \citealp{2019MNRAS.485.1651L}). This feature was first spotted in the Centaurus cluster \citep{2002MNRAS.331..273S} and later found in other galaxy groups and clusters (e.g., \citealt{2015MNRAS.447..417P}). Initially, the Fe drop was suspected to be an artifact arising from the ``Fe bias'' (\citealp{1998MNRAS.296..977B}, \citealp{2018SSRv..214..129M}, \citealp{2000MNRAS.311..176B}), introduced by the oversimplified assumption of the single-temperature plasma. However, even considering multi-temperature models and deeper exposures, the Fe drop remains evident in several clusters \citep{2019A&A...623A..17L}. Resonance scattering is also proposed as a potential cause for the underestimation of central Fe abundance \citep{2006MNRAS.370...63S}, but subsequent studies suggest its influence on the observed Fe drop is limited \citep{2009MNRAS.398...23W, 2011MNRAS.411..411C, 2017ApJ...848...26G}. Similar central abundance drop is also reported on other elements \citep{2015MNRAS.447..417P}.

Since Fe drop is usually accompanied by X-ray cavities, it is possibly associated with AGN activity \citep{2015MNRAS.447..417P}. Simulations indicate that strong mechanical feedback of central AGN may transfer a CC to a non-cool-core (NCC), also flattening its central Fe peak profile \citep{2010ApJ...717..937G}. An alternative explanation is the ``dust depletion scenario" \citep{2013MNRAS.433.3290P}, which proposes that a fraction of metals condense onto dust grains. These dust grains are then entrained by AGN-driven outflows, heated, and sputtered back into the hot ICM at several kiloparsecs from the center. This scenario could be verified via the abundance profile of noble gas elements (e.g., Ne and Ar) because they are not easily deposited into dust grains. Though some have found a trend of Ne/Fe increase to the center, the Ne abundances derived by CCD observations have very large uncertainties (e.g., \citealp{2019MNRAS.485.1651L}).

Among all targets spotted to have a possible Fe drop feature, MACS J1931.8-2634 (hereafter MACS J1931) has the most extended Fe drop detection to about $100~{\rm kpc}$ \citep{2011MNRAS.411.1641E}. MACS J1931 is a massive ($M_{500}\sim4.96\times10^{14}~M_\odot$, \citealp{2021ApJS..253....3H}) CC cluster at redshift $z=0.352$ \citep{2019A&A...632A..36C}, corresponding to the scaling of about $4.96~{\rm kpc/arcsec}$.  $R_{500}\sim0.94~{\rm Mpc}$ could be derived from $R_{500}=\left(\frac{3M_{500}}{4\pi\times500\rho_{\rm crit}(z)}\right)^{1/3}$, where $\rho_{\rm crit}$ is calculated in $\Lambda$CDM cosmology with $H_0=70~{\rm km/Mpc/s}$, $\Omega_m=0.3$, $\Omega_\Lambda=0.7$. 
The brightest cluster galaxy (BCG) in this cluster has an SFR of approximately $250~M_\odot~{\rm yr^{-1}}$ and hosts an X-ray bright AGN \citep{2017ApJ...846..103F}, possibly an ideal candidate to study the large-scale feedback. Radio and far-infrared observations reveal $(1.9\pm0.3)\times10^{10}~M_\odot$ of molecular gas \citep{2019ApJ...879..103F}, making it one of the largest known reservoirs of cold gas in a cluster core. Detection of dust filaments hints possible dust involvement in the metal distribution. 
In this letter, we derive both 1D and 2D Fe distribution from new XMM-Newton observations and Chandra archival data, trying to confirm the extended Fe drop feature and find possible correlations with AGN activity. We also measure the abundance ratio of Ne/Fe in the central region with high-resolution spectroscopy to find if dust depletion is involved.

\section{Data reduction and spectral analysis}
\subsection{Observations and data reduction}
\label{sct:obs_dr}

\begin{table}[h!]
\renewcommand\arraystretch{1.3}
\caption{X-ray observation logs}           
\label{table:obs-log}      
\centering          
\begin{tabular}{ c c c c }
\hline\hline       
 Obsid & Date & Instrument & Net exp (ks)\\
 \hline
 \hline
\multicolumn{4}{c}{Chandra}\\
\hline                    
 3282 & 2002-10-20  & ACIS-I & 13.6\\
\hline
 9382 & 2008-08-21  & ACIS-I & 98.9\\
\hline
\hline
\multicolumn{4}{c}{XMM-Newton}\\
\hline
 \multirow{4}{*}{0693180101} & \multirow{4}{*}{2012-10-18}  & MOS1 & 38.8 \\
 & & MOS2 & 39.2\\
 & & RGS1 & 29.5\\
 & & RGS2 & 32.6\\
\hline
 \multirow{4}{*}{0920170101} & \multirow{4}{*}{2023-10-27}  & MOS1 & 78.5\\
 & & MOS2 & 84.8\\
 & & RGS1 & 71.8\\
 & & RGS2 & 69.4\\
\hline
 \multirow{4}{*}{0920170201} & \multirow{4}{*}{2023-10-29} & MOS1 & 60.6\\
 & & MOS2 & 49.1\\
 & & RGS1 & 76.3\\
 & & RGS2 & 76.1\\
\hline
\end{tabular}
\end{table}
We reduced all the archived XMM-Newton observations of MACS J1931.8-2634, including two new observations obtained in Cycle 22 (PI: J. Mao).
The European Photon Imaging Camera (EPIC) and Reflection Grating Spectrometer (RGS; \citealp{2001A&A...365L...7D}) spectra were extracted with XMM-Newton Science Analysis System (SAS) v22.1.0 following \citet{mao2023xmmnewtonreflectiongratingspectrometer}. Due to the systematic differences between EPIC/MOS \citep{2001A&A...365L..27T} and EPIC/PN \citep{2001A&A...365L..18S}, introduced by under- or over-estimation of PN particle background (see: ESAS cookbook\footnote{https://heasarc.gsfc.nasa.gov/docs/xmm/esas/cookbook/}, section 1.3), we excluded PN data from our analysis. Unfortunately, the EPIC-MOS data are highly influenced by flares. About 20\% to 40\% of the time is contaminated by flares, which were excluded from good time intervals (GTIs). For comparison, we also re-analyzed the archival Chandra ACIS observations using CIAO v4.15 software package \citep{2006SPIE.6270E..1VF} and CALDB v4.10.7. The observation logs and exposure time of cleaned data we used are recorded in Table~\ref{table:obs-log}. 

We extracted RGS spectra from $\sim 0.8~\rm arcmin$ wide slit (including 90\% PSF) along the cross-dispersion direction. For joint fitting with RGS spectra, MOS spectra were extracted from a circular region with a radius equal to the half-width of the RGS extraction slit ($0.4~{\rm arcmin}$), which leads to a slightly different aperture \citep{2021ApJ...918L..17M}.

To construct radial profiles, CCD spectra were extracted from a series of concentric annuli centered on the AGN. The inner and outer radii of these annuli are 0 (for MOS)/7.3 (for ACIS), 50, 100, 150, 200, 300, and 400 kpc. For Chandra, a central region of approximately 7.3 kpc was excluded to mitigate contamination from the AGN point source. For MOS instead, the larger PSF makes it difficult to effectively exclude AGN contamination through region selection alone; therefore, we included an AGN component in the spectral modeling referred to \citet{2011MNRAS.411.1641E}. Local background spectra were extracted from the annulus centered on the BCG, with radii from $R_{500}$ to $1.5~R_{500}$, after removing detected point sources.  

In addition to the radial profiles, we generated two-dimensional maps using the weighted Voronoi tessellation (WVT) method \citep{2006MNRAS.368..497D}. The WVT bins were constructed from the EPIC/MOS images, with a target signal-to-noise ratio of $\rm S/N \geq 70$ in each bin. The same set of spatial regions was then used for spectral extraction. For most regions, we fitted the EPIC/MOS spectra. However, for the five innermost regions, the broader MOS PSF makes it difficult to account for the effect of central AGN. We therefore extracted and fitted Chandra/ACIS spectra for these regions using the same WVT boundaries, while excluding the central point source from the ACIS data. The resulting ACIS spectra in these central regions have $\rm S/N \geq 50$. This procedure allows us to retain the MOS-defined spatial binning while reducing AGN contamination in the cluster-core spectra.

\begin{figure}
  \resizebox{\hsize}{!}{\includegraphics{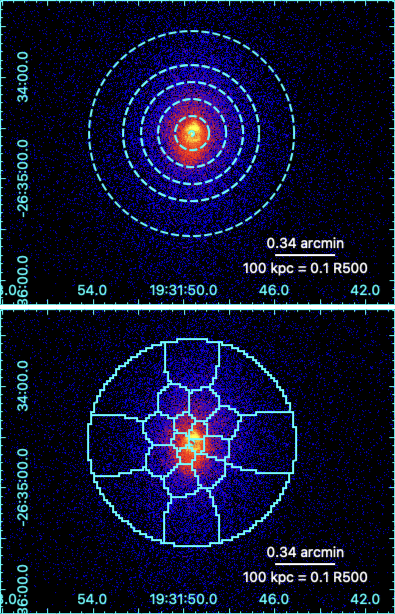}}
  \caption{Region of interests for generating 1D profile (left) and 2D maps (right) overplotted on Chandra combined image. For both regions, the largest circle corresponds to $300~{\rm kpc}$.}
  \label{RegionofInterest}
\end{figure}

\section{Spectral analysis}

All spectral fitting was performed using SPEX v3.07.03  \citep{kaastra_2020_3939056} with C-statistics \citep{2017A&A...605A..51K}. The proto-solar abundance table of \citet{2009LanB...4B..712L} was used throughout this work. 

The hot ICM was modeled with an absorbed collisional ionisation equilibrium (cie) model, both single-temperature (1-T) and multi-temperature (multi-T) situations are considered. For the multi-T scenario, the differential emission measure distribution follows a log-normal function (GDEM; \citealt{2006A&A...452..397D}):
\begin{equation}
    Y(x)=\frac{Y_0}{\sigma\sqrt{2\pi}} \exp{\left( -\frac{(x-x_0)^2}{2\sigma^2}\right)}
\end{equation}
where $Y$ and $Y_0$ represent the emission measures, $x=\log_{10}T,~x_0=\log_{10}(T_0)$, and $T_0$ is peak temperature in units of keV. For the outer regions observed by Chandra, the signal-to-noise ratio was insufficient to constrain multi-T models; in these cases, we adopted the results from 1-T fits.

\begin{figure}
  \resizebox{\hsize}{!}{\includegraphics{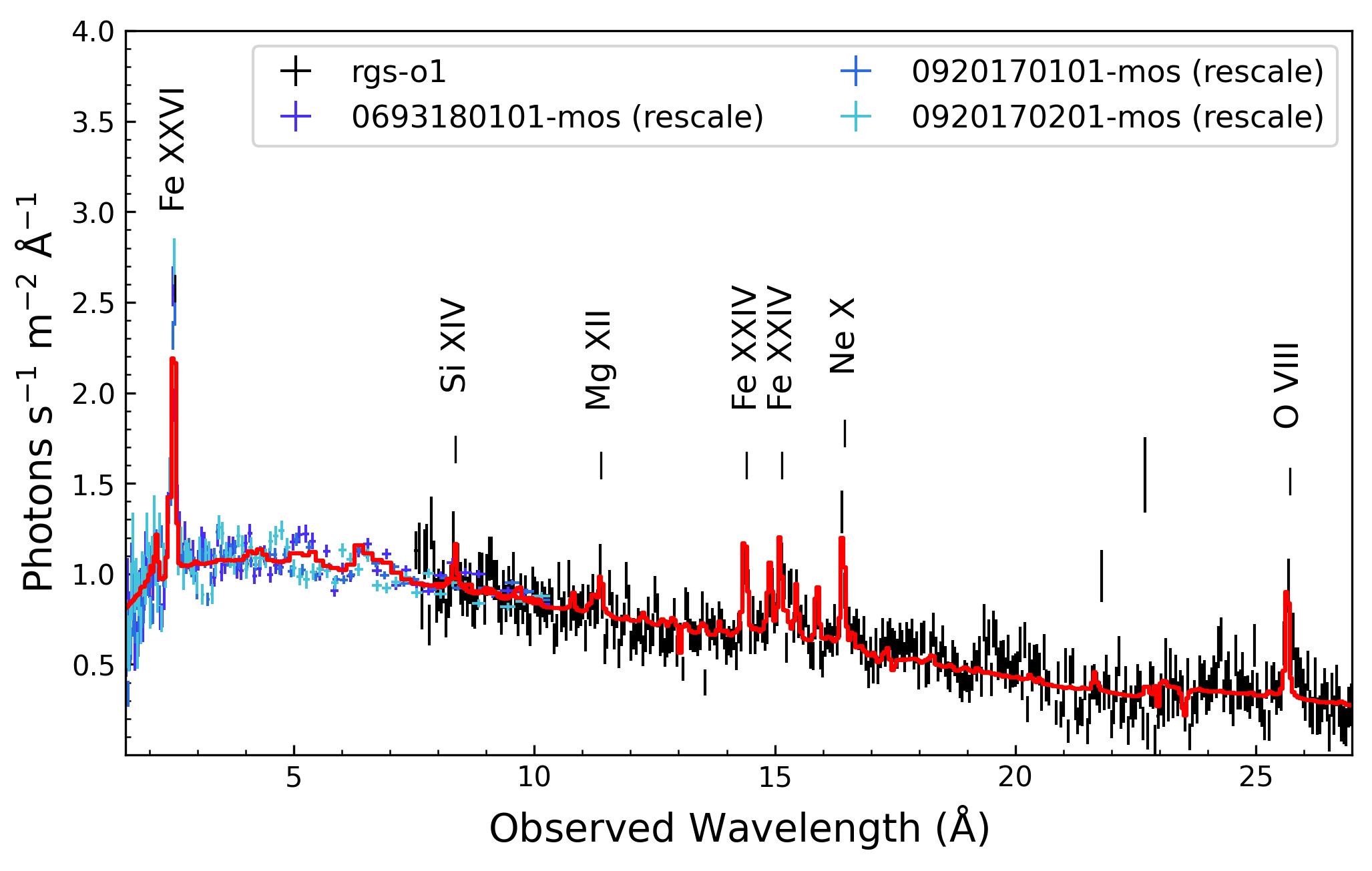}}
  \caption{The best-fit model (multi-T) to the X-ray spectrum of MACS J1931 observed by XMM-Newton. The high-resolution RGS spectra are shown in black, rebinned by a factor of 5 for plotting only. The EPIC/MOS spectra are shown in blue, rescaled by a factor of 1.72.}
  \label{RGS spectra}
\end{figure}

The RGS spectra provide the strongest constraints on the O, Ne, and Fe-L line complexes, but their relatively narrow bandpass alone does not tightly constrain the broad-band continuum level and temperature structure of the thermal plasma. We therefore included the EPIC/MOS spectra to improve the constraints on these quantities, following a similar approach to that adopted by \citet{2021ApJ...918L..17M}. We simultaneously fitted the RGS spectra in $7.5-27~\rm\AA$ and MOS spectra in $1.2-8.5~\rm keV$ ($\sim 1.5-10~\rm\AA$), as shown in Figure~\ref{RGS spectra}. This choice of energy band for MOS prevents the MOS data from dominating the statistics in the soft X-ray band, while retaining a small spectral overlap to tune the scaling between the two instruments. RGS spectra were binned by a factor of 2, while EPIC/MOS were optimally binned \citep{2016A&A...587A.151K}. A scaling factor of 1.72 was applied to the EPIC/MOS spectra relative to the RGS to account for differences in instrument normalization and extraction apertures. In these fits, we left the abundances of O, Ne, Mg, Si, and Fe free. The best-fit results are reported in Table~\ref{tab:RGS-result}.

The MOS (0.5–8.5 keV) and ACIS  (0.5–7.5 keV) spectra were fitted independently, since they suffer from different instrumental effects. All spectra were optimally binned, and the local background was subtracted directly. We set free only the Fe abundances. Due to the limited spectral resolution of CCD spectra, emission lines from elements other than Fe are weak, blended, or not sufficiently resolved in individual spatial bins, thus fixed to the proto-solar values. Since the central region extracted from MOS would be affected by the AGN, we included an additional absorbed power-law component. According to \citet{2011MNRAS.411.1641E}, the photon index is set to 1.70, intrinsic $n_{\rm H}$ set to $7.1\times10^{21}~{\rm cm^{-2}}$, and the flux is $1.95\times10^{-13}~\rm {erg~cm^{-2}s^{-1}}$.

\section{Results and discussions}
\subsection{Abundance ratios from RGS}
One explanation for the phenomenon of Fe drop is the dust depletion scenario proposed in \citet{2013MNRAS.433.3290P}. Some of the metal-rich gas injected by Type Ia SNe or stellar feedback may remain clumped, where iron is easily depleted onto grains. The buoyant bubbles generated in the radio mode of AGN activity will possibly drag out these clumps. While the metal-rich dust grains are moving outward, they are heated by the surrounding hot ICM and generally eroded by sputtering. The iron is then returned to the hot ICM and contributes to the high metallicity away from the center. Since noble gas elements such as Ne are hard to be depleted into dust, this scenario could be verified by measuring the Ne/Fe abundance ratio. However, the Ne/Fe ratio is also influenced by the Type Ia to core-collapse supernovae ratio. We therefore also measured the Ne/O abundance ratio as a reference, since O is also primarily produced by core-collapse supernovae.

\begin{table}[h!]
\renewcommand\arraystretch{1.3}
\caption{Joint fitting result of RGS and MOS for central 100 kpc.}
\centering
\begin{tabular}{c|c|c}
\hline\hline
Instrument & \multicolumn{2}{c}{RGS+MOS}\\
\hline
Model & 1-T & multi-T \\
\hline
$T/\rm keV$ & $5.14^{+0.06}_{-0.06}$ & $5.30^{+0.21}_{-0.08}$ \\
sig & - & $0.41^{+0.02}_{-0.02}$\\
$Z_{\rm O}/Z_\odot$ & $0.72^{+0.13}_{-0.12}$ & $0.51^{+0.09}_{-0.09}$ \\
$Z_{\rm Ne}/Z_\odot$ & $1.34^{+0.23}_{-0.60}$ & $0.67^{+0.16}_{-0.15}$ \\
$Z_{\rm Mg}/Z_\odot$ & $0.56^{+0.30}_{-0.28}$ & $0.47^{+0.22}_{-0.20}$ \\
$Z_{\rm Si}/Z_\odot$ & $0.24^{+0.11}_{-0.10}$ & $0.36^{+0.08}_{-0.08}$ \\
$Z_{\rm Fe}/Z_\odot$ & $0.59^{+0.02}_{-0.02}$ & $0.65^{+0.03}_{-0.03}$ \\
\hline
C-stat & 1407 & 1322 \\
C-exp & $1248\pm50$ & $1248\pm50$ \\
\hline
\end{tabular}
\label{tab:RGS-result}
\end{table}

The joint fitting result of central $100~{\rm kpc}$ spectra by RGS and MOS is shown in Table~\ref{tab:RGS-result}. The single temperature and multi-temperature models yield significantly different abundances, especially for Ne. The 1-T model gives a supersolar Ne/Fe ratio ($>2$), while the multi-T model yields ${\rm Ne/Fe}=1.03^{+0.25}_{-0.23}$, ${\rm O/Fe}=0.79^{+0.15}_{-0.14}$ and ${\rm Ne/O}=1.30^{+0.38}_{-0.37}$. Our multi-T result for Ne/Fe is consistent with solar ratio, similar to the measurements in the Perseus cluster core \citep{2019MNRAS.483.1701S}. Though Ne/O ratio is slightly above solar, it is consistent with the ratio measured in the hot atmosphere of galaxies (e.g., \citealp{2021ApJ...918L..17M}). These abundance ratios do not show a significant enhancement of Ne relative to Fe, and therefore do not support the dust depletion scenario as the origin of a central Fe deficit.

The high star formation rate of the BCG also raises the question of whether recent core-collapse supernova enrichment should enhance the $\alpha$-element abundance relative to Fe in the hot ICM. Since O and Ne are mainly produced by massive stars and core-collapse supernovae, efficient mixing of newly synthesized metals into the X-ray-emitting gas could lead to elevated O/Fe and Ne/Fe ratios. For example, the luminous infrared galaxy Arp 299 shows an O/Fe of around 2.5 and Ne/Fe of around 2.9 \citep{2021ApJ...918L..17M}. In contrast, MACS J1931 does not show such an enhancement in its central hot ICM. The O/Fe ratio is instead close to the CHEERS average abundance pattern of nearby cool-core systems, where ${\rm O/Fe}=0.817\pm0.175$ and $\rm{Ne/Fe}=0.724\pm0.133$ \citep{2018MNRAS.480L..95M}. Therefore, although the BCG in MACS J1931 is undergoing intense star formation, the present RGS abundance ratios do not provide evidence that recent core-collapse supernova products dominate the chemical pattern of the X-ray-emitting ICM on the central $ 100~\rm kpc$ scale. However, this may also be related to the dispersive nature of the RGS, which leads to spatial mixing of emission from different radii and may dilute centrally concentrated features \citep{2001A&A...365L...7D}.

\subsection{Radial profile}
\label{sec:radial profile}
\begin{figure*}
\sidecaption
  \includegraphics[width=12cm]{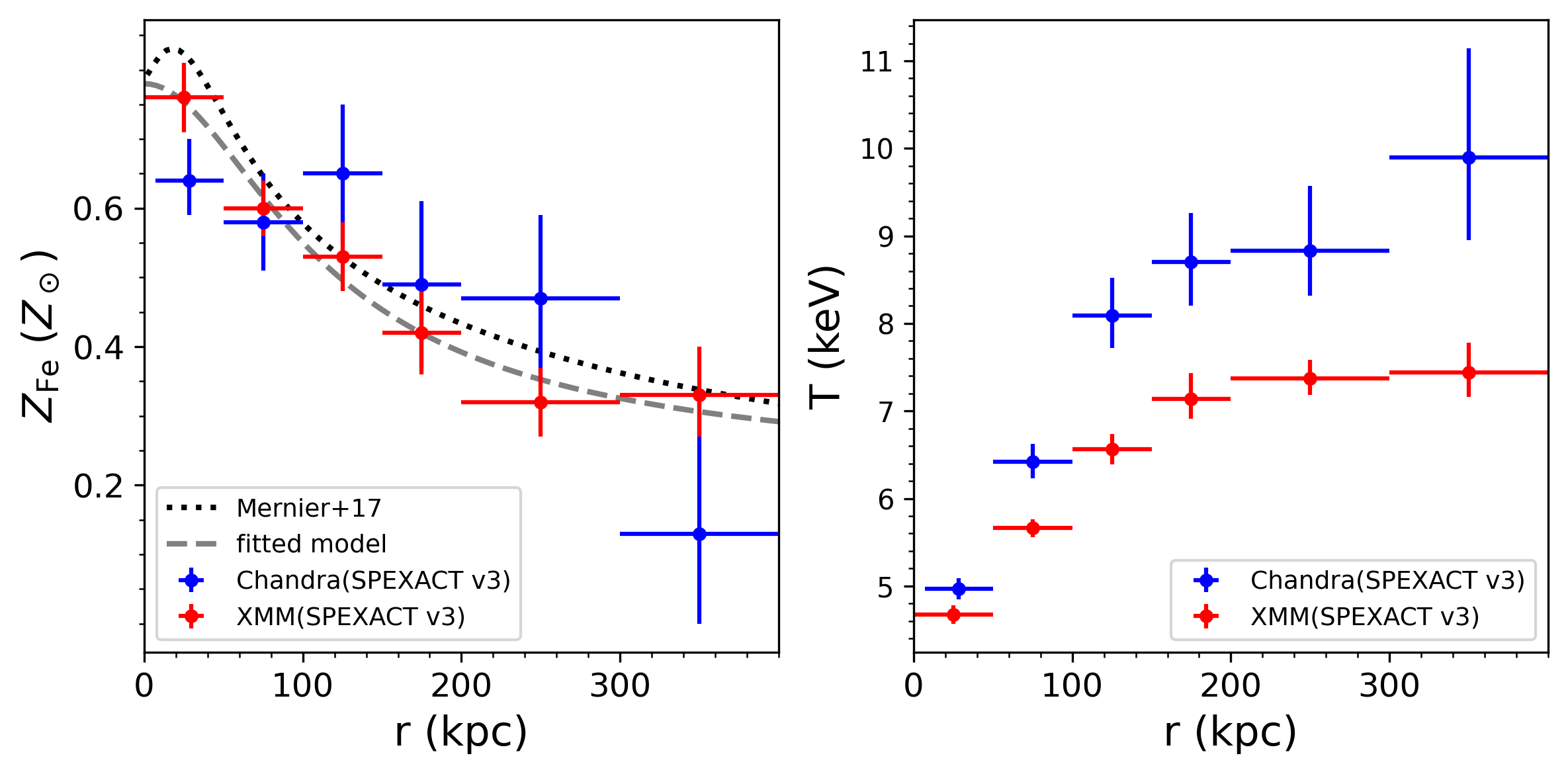}
     \caption{Radial profile of $Z_{\rm Fe}$ and temperature obtained from both Chandra/ACIS (shown in blue crosses) and XMM/EPIC-MOS (red crosses). All the spectra are fitted with SPEXACT v3. In the left panel, the gray line represents the best-fit model to the XMM-derived $Z_{\rm Fe}$ profile. The black dot line a stacking profile of 44 nearby cool-core galaxy groups and clusters (CHEERS) by \citet{2017A&A...603A..80M}.}
     \label{fig:radial profile}
\end{figure*}

The radial profiles of temperature and Fe abundance derived from MOS and ACIS are shown in Figure~\ref{fig:radial profile}. A systematic offset in temperature is evident, with Chandra values being consistently higher. \citet{2015A&A...575A..30S} have compared the temperature estimation of 46 galaxy clusters by the two observatories, and found that the temperature given by Chandra is generally higher than that given by XMM-Newton. The discrepancy increases with higher temperature, up to 20\%, which is similar to the result in our work. Also, recent work shows the deviation of Chandra-derived temperature, while the temperature by other X-ray instruments remains consistent \citep{2025PASJ...77S.254S}. 

Due to the degeneracy of temperature and metallicity, the Fe abundance profiles derived from the two observatories differ, particularly at larger radii. While our ACIS results show a hint of an Fe drop within 100 kpc, consistent with \citet{2011MNRAS.411.1641E}, this feature is absent in the MOS profile. We also examined the deprojected MOS profile, which shows no central Fe drop.

The MOS Fe abundance can be well described by a simple $\beta$ model with a constant plateau, representing the central peak and background metallicity \citep{2020A&A...637A..58L}:
\begin{equation}
Z_{\rm Fe,1D} = Z_{\rm peak}\times[1+(r/r_0)^2]^{-\alpha}+Z_{\rm plateau}
\label{eq:Z_1D}
\end{equation}
The best-fit value gives: $Z_{\rm peak}=0.57~{Z_\odot}$, $r_0 = 90.0~{\rm kpc}$, $\alpha=0.64$, and $Z_{\rm plateau}=0.21~{Z_\odot}$. No additional ``drop'' component is required. Compared with the radial profile of the stacked 44 nearby galaxy clusters and groups (CHEERS sample) obtained in \citet{2017A&A...603A..80M},
\begin{equation}
    Z_{\rm Fe} = 0.21(r+0.021)^{-0.48}-6.54\exp{\left(-\frac{(r+0.0816)^2}{0.0027}\right)}
\end{equation}
MACS J1931 also shows consistency within errors.

\subsection{2D maps}
The disruption of AGN activity or other dynamical processes is not expected to be isotropic, nor is the distribution of metals. To explore the complex, asymmetric metal distribution that may be missed in purely radial analysis, we constructed two-dimensional maps of Fe abundance and temperature using the methods described in Section~\ref{sct:obs_dr}.

\begin{figure*}
\sidecaption
  \includegraphics[width=12cm]{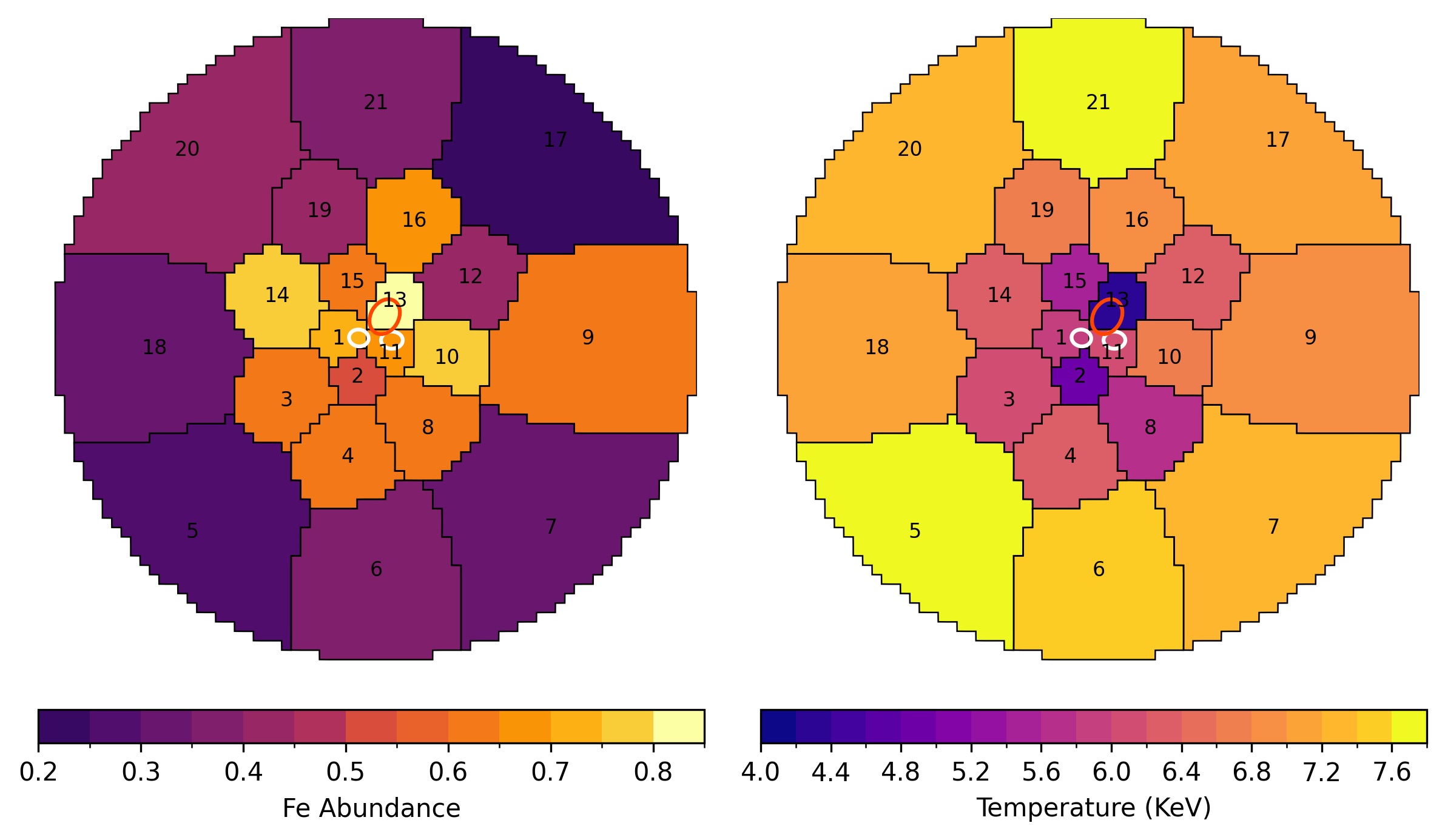}
     \caption{Voronoi binning map of Fe abundance and temperature. Spectral results are obtained from Chandra/ACIS for the five central regions (1, 2, 11, 13, and 15), and from XMM/EPIC-MOS for the remaining regions. The maps show the spectral fitting result by multi-temperature models, with the temperature taking the peak value. The white ellipses outline the location of X-ray cavities. The red ellipse corresponds to the location of optical filament observed by MUSE.}
     \label{fig:vtbin}
\end{figure*}

As shown in the weighted Voronoi binning map (Figure~\ref{fig:vtbin}), the most Fe-rich region (region 13, with $Z_{\rm Fe}=0.84^{+0.09}_{-0.09}$) coincides with the brightest X-ray emission and is slightly offset from the central AGN. This region also aligns spatially with optical line emission traced by MUSE contours \citep{2021A&A...649A..23C}. Several other regions, including region 14, 1, 10, and 9, also show relatively high Fe abundances compared with nearby bins. These enriched regions are located roughly along the axis of the X-ray cavities (indicated by white circles in the figure), although they extend beyond the immediate cavity scale and reach distances of $100-300~{\rm kpc}$.

\begin{figure*}
\sidecaption
  \includegraphics[width=12cm]{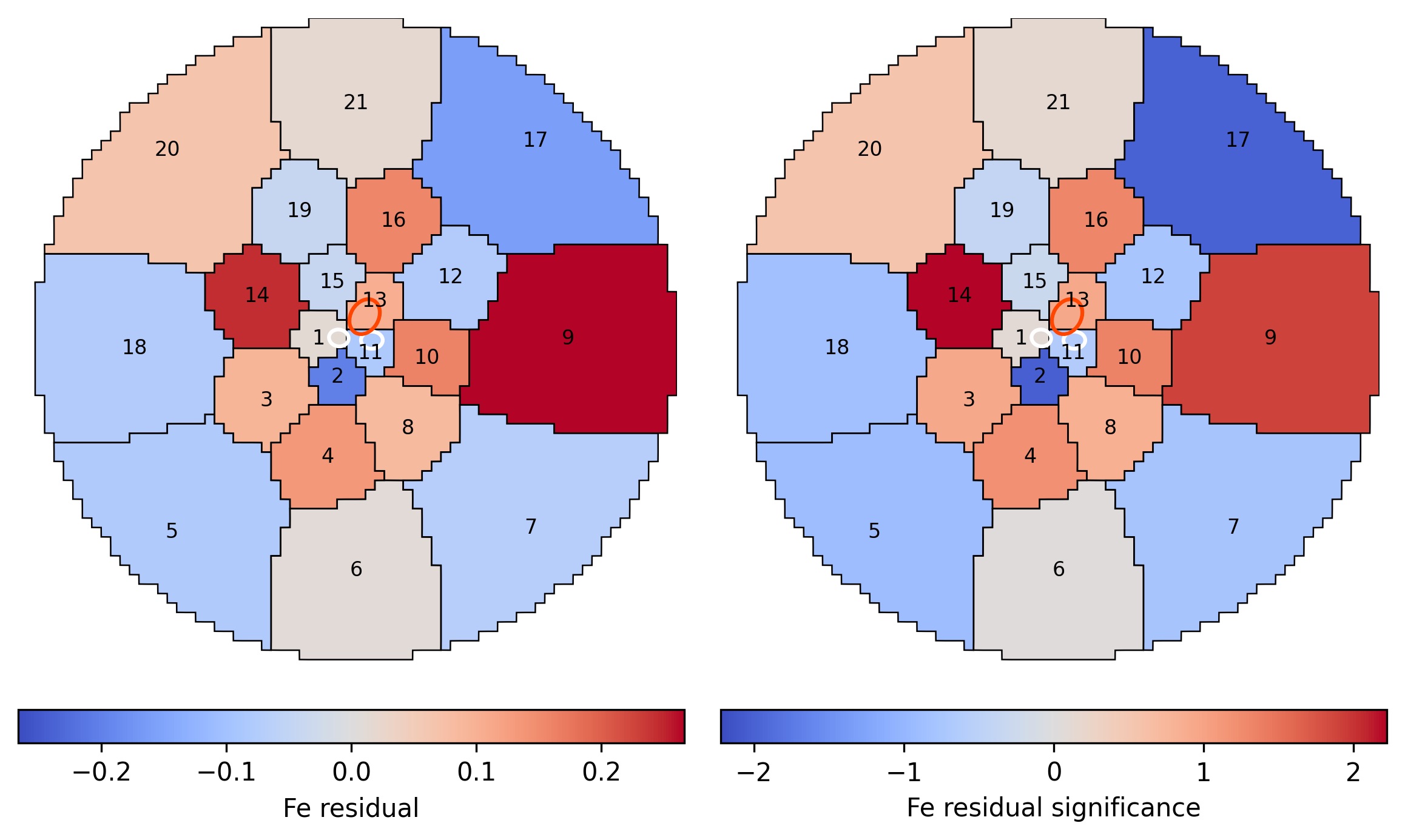}
     \caption{Left: Fe residual map, defined as the difference between the measured 2D Fe abundance and the best-fit radial Fe abundance profile in each Voronoi bin, following Eq.(\ref{eq:residual}). Right: Fe residual significance map, defined as the Fe residual normalized by the $1-\sigma$ statistical uncertainty of the Fe abundance in the corresponding bin, following Eq.(\ref{eq:residualsig}).}
     \label{fig:Fe-residual}
\end{figure*}

To assess whether this apparent asymmetry is driven only by the radial abundance gradient, we first constructed a Fe residual map (Figure~\ref{fig:Fe-residual}, top) relative to the azimuthally averaged 1D Fe profile. This removes the dominant radial trend and highlights azimuthal deviations from the average profile. For each Voronoi bin, we evaluated the best-fit radial abundance model at the radius of the bin center and defined the Fe residual as
\begin{equation}
\Delta Z_{\rm Fe}=Z_{\rm Fe,2D}-Z_{\rm Fe,1D}(r)
\label{eq:residual}
\end{equation}
where $Z_{\rm Fe,2D}$ is the Fe abundance measured in each bin and $Z_{\rm Fe,1D}(r)$ is the expected abundance calculated from the best-fit $\beta$ model, see Eq.(\ref{eq:Z_1D}). We then constructed a Fe residual significance map (Figure~\ref{fig:Fe-residual}, bottom) by normalizing the same residuals by the statistical uncertainty,
\begin{equation}
S=\frac{\Delta Z_{\rm Fe}}{\sigma_{Z_{\rm Fe}}}
\label{eq:residualsig}
\end{equation}
where $\sigma_{Z_{\rm Fe}}$ is the 1-$\sigma$ uncertainty of the Fe abundance in that bin. Thus, the Fe residual map shows the absolute amplitude and sign of the deviation from the radial model, while the Fe residual significance map provides the corresponding uncertainty-weighted view.

The two maps show broadly similar spatial patterns, but the significance map provides a more direct reference to assess which deviations are robust relative to their statistical uncertainties. The positive residuals are not confined to the central peak. The strongest positive residual appears in region 14, reaching approximately 2-$\sigma$ level. Region 10 also shows one of the relatively high positive residuals, although its individual significance is more modest, at about the 1-$\sigma$ level. However, this result should be interpreted with caution, as this region is located at a larger radius and covers a relatively large area. Region 10 is large and located at a greater radius, where the lower surface brightness increases sensitivity to the XMM-Newton background treatment (e.g., \citealp{2005ApJ...629..172N}). The large bin may also mix gas with different temperatures and abundances, which can bias Fe abundance measurements in multi-temperature plasma (e.g., \citealp{2018SSRv..214..129M}). The remaining residuals are not individually significant. Nevertheless, the residual maps as a whole suggest a large-scale asymmetry in the Fe distribution, with positive residuals preferentially distributed along the cavity axis.

Previous studies of cool-core clusters such as Hydra-A \citep{2009ApJ...707L..69K}, MS 0735.6+7421 \citep{2014MNRAS.442.3192V}, and M87 \citep{2008A&A...482...97S} have reported that high-metallicity gas is often elongated along the direction of AGN jets. In our case, however, the spatial mismatch between the cavity scale and the extended Fe enhancement makes it difficult to attribute the enrichment pattern solely to ongoing AGN-driven transport. Whether this large-scale asymmetry reflects past AGN episodes, merger-induced mixing, or a combination of processes remains an open question.

Nonetheless, the 2D map clearly reveals that the Fe distribution is asymmetric, which is hard to spot from the radial profile alone. Such azimuthal variations may be common in other clusters as well, suggesting that the full complexity of chemical enrichment is best captured through spatially resolved mapping. Inspired by quantitative methods used to characterize cluster dynamical states from X-ray surface brightness, similar morphological statistics could also be applied directly to metallicity distribution maps. Parameters quantifying the asymmetry, concentration, and clumpiness of metal distributions could provide powerful, complementary diagnostics. Future studies with deeper, high-resolution X-ray spectroscopy (e.g., NewATHENA; \citealt{2025NatAs...9...36C}, HUBS; \citealt{2020JLTP..199..502C}) will be crucial to construct detailed, multi-element abundance maps for a larger sample of clusters.  Applying such chemical morphology indices across an increasing population of clusters will help systematically characterize the ``ecological diversity'' of the ICM. Also, by linking the spatial distribution of metals to the thermodynamic morphology, we could better understand how feedback processes—such as AGN outbursts, mergers, or gas sloshing—operate and leave their distinct imprints on the chemical and dynamical state of clusters.

\begin{acknowledgements}
The authors thank the referee for the careful review and valuable suggestions that helped improve the paper. Q. Z. gratefully acknowledges useful discussions and consultations about XMM-Newton data reduction with Yi-Heng Chi, Jiejia Liu, Qingling Ni, and also thanks L\'{y}dia \v{S}tofanov\'{a} for helpful discussions on SPEX modeling. J. M. acknowledges support from the Tsinghua Dushi Program 53121200125. P. Z. thanks the support by the National Natural Science Foundation of China with grant 12273010 and the Fundamental Research Funds for the Central Universities No. KG202502. A reproduction package is available in \href{https://sandbox.zenodo.org/records/515798}{10.5072/zenodo.515798}.
\end{acknowledgements}

\bibliographystyle{aa}
\bibliography{MJ1931}

\end{document}